\documentclass[prl,twocolumn,showkeys]{revtex4}
\usepackage{amsmath,bm}
\usepackage{fleqn}

\newcommand{\B}{\mathbf{B}}
\renewcommand{\d}{\mathbf{d}}
\newcommand{\E}{\mathbf{E}}
\newcommand{\e}{\mathbf{e}}

\newcommand{\p}{\mathbf{p}}
\renewcommand{\r}{\mathbf{r}}
\newcommand{\s}{\mathbf{s}}
\newcommand{\x}{\mathbf{x}}
\newcommand{\obe}{\overline e}
\newcommand{\obF}{\overline F}
\newcommand{\obh}{\overline h}
\newcommand{\obm}{\overline m}
\newcommand{\obS}{\overline S}
\newcommand{\obp}{\overline p}
\newcommand{\obu}{\overline u}
\newcommand{\obv}{\overline v}

\renewcommand{\AA}{\hbox{$\mathring{\textrm{A}}$}}

\newcommand{\bsig}{\bm{\sigma}}
\newcommand{\bmu}{\bm{\mu}}
\newcommand{\bnabla}{\bm{\nabla}}

\newcommand{\bdot}{\bm{\cdot}}

\begin{document}

\title{Reading the Electron Clock}

\author{David Hestenes}
\affiliation{Department of Physics, Arizona State University, Tempe, Arizona 85287-1504}
\email{hestenes@asu.edu}
\homepage{http://modelingnts.la.asu.edu/}

\begin{abstract}
If electron zitterbewegung is a real effect, it should generate an electric dipole field oscillating with the zitterbewegung frequency $2m_ec^2/\hbar$.  The possibility of detecting it as a resonance in electron channeling is analyzed.
\end{abstract}


\maketitle

\section{Introduction}
In his seminal contribution to quantum mechanics, Louis de Broglie \cite{Broglie23}
conjectured that the electron has an internal clock oscillating with frequency
$\nu_B=m_ec^2/h$ determined by its rest mass $m_e$. He then deduced that the frequency of
a moving electron observed in a laboratory will be $\nu_L=\nu_B/\gamma$, where $\gamma$ is
the relativistic time dilation factor. However, this idea of a particle clock was soon
forgotten after the creation of wave mechanics, wherein the \textit{de Broglie frequency}
$\,\,\omega_B=2\pi\nu_B$ is interpreted exclusively as the frequency of a wave.
Besides, detecting ticks in an electron clock seemed out of
the question because of the ultra high frequency $\omega_B=7.7634\times
10^{20}\text{s}^{-1}$. The possibility of direct detection as a resonance in electron
channeling was recognized only recently and tested with positive results in an exploratory
experiment \cite{Gouanere05}.

When a beam of electrons is channeled along a crystal axis, each electron is subjected to
periodic impulses from atoms along the axis.
When the energy of the beam is adjusted so the period of impulses matches the period of the
electron's clock, a resonant interaction should occur if the clock interacts with the atoms
in some way. The resonant energy is easily calculated from de Broglie's hypothesis.
The distance traversed during a clock period is $d=c\beta/\nu_L=hp/(m_ec)^2$.
For the silicon crystal used in the experiment, the interatomic distance along the $<110>$
direction is $d=3.84\,\AA$, which implies a resonant momentum $p=80.874$ MeV/c. A dip in the
transmission rate of the channeled electron beam was detected at 81.1 MeV/c, just 0.28\% from the
expected resonance and within the estimated 0.3\% limit on calibration error.

This result may seem doubtful, as it was not anticipated nor is it likely to be explained by
standard quantum mechanics, despite extensive work on channeling theory.
However, possibilities for an explanation are suggested by oscillations in solutions of
the Dirac equation called \textit{zitterbewegung} by Schroedinger \cite{Schr30}.
Many physicists, including Dirac himself \cite{Dirac58}, have suggested that zitterbewegung
describes local oscillations in electron position that account for the electron's spin
and magnetic moment. At first, Dirac argued for the existence of an electric dipole
moment fluctuating with zitterbewegung frequency $\omega_Z=2\omega_B$ \cite{Dirac28}, but he
soon retracted that suggestion. However, the electron clock as a rotating dipole may be just
what we need to explain the interaction presumed in the channeling experiment, and it
happens that $\omega_Z$ can account for the data at least as well as $\omega_B$. Indeed, the
two frequencies are intimately related. For a free particle, $\omega_B$ appears in the phase
of a Dirac wave function, so observables, which are bilinear functions of the wave function,
exhibit the frequency $\omega_Z$.

Guided by analysis of Dirac theory, a new particle model of the electron has recently been
developed wherein spin and zitterbewegung are complementary features of electron
kinematics, and interaction with arbitrary electromagnetic fields is included \cite{Hest08}.
The present paper employs that model to explain the putative electron
clock resonance in channeling, and extend it with quantitative
predictions for shifts in zitterbewegung frequency that can be tested with more precise
experiments. If confirmed, the channeling experiments will thus provide the first direct
evidence that zitterbewegung is a real physical effect. Implications for quantum
mechanics are likely to be profound \cite{Hest08}.

\section{ The Zitter Model}

The word ``zitterbewegung'' is quite a mouthful, so let us shorten it to \textit{zitter}!
In developing the model of a point particle with zitter, geometric algebra has played an
essential role, and it greatly facilitates solving the coupled system of equations in the
model as well as comparison with Dirac theory \cite{Hest08}. However, for convenience of
readers unfamiliar with geometric algebra, a synopsis of the model in standard
tensor form is presented here (using natural units until calculations are in order).

We model the electron as a point particle in spacetime with a lightlike history
$z_\mu=z_\mu(\tau)$, so its \textit{velocity}
\begin{equation}
u_\mu\equiv \frac{dz_\mu}{d\tau}\qquad\hbox{satisfies}\qquad u_\mu u^\mu=0,\label{1}
\end{equation}
where $\tau$ is an intrinsic time determined by the equations of motion (call it \textit{electron time}). The particle's \textit{momentum} is a timelike vector $p_\mu$, and it determines a
dynamical mass $m$ defined by
\begin{equation}
 p^\mu p_\mu=m^2>0, \qquad\hbox{and}\qquad m=p^\mu u_\mu.\label{2}
\end{equation}
The particle also has intrinsic angular momentum (or \textit{spin}) characterized by a tensor
$ S_{\mu\nu}=-S_{\nu\mu}$ with the properties
\begin{equation}
 u^\mu S_{\mu\nu}=0\qquad\hbox{and}\qquad
S_\mu\,^\alpha\, S_{\alpha\nu}=0.\label{3}
\end{equation}
The particle is charged, so it interacts with any external electromagnetic field
$F_{\mu\nu}=\partial_\mu A_\nu-\partial_\nu A_\mu$.

Particle motion is determined by a system of coupled equations for velocity,
momentum and spin:
\begin{equation}
 \frac{du_\mu}{d\tau}  =\frac{4}{\hbar^2} p^\nu S_{\nu\mu}+
\frac{q}{m_e}F_{\mu\nu}u^\nu,\label{4}
\end{equation}
\begin{equation}
 \frac{dp_\mu}{d\tau}=qF_{\mu\nu}u^\nu
+\frac{q}{2m_e}S^{\beta\alpha}\partial_\mu F_{\alpha\beta},\label{5}
\end{equation}
\begin{equation}
  \frac{dS_{\mu\nu}}{d\tau}=u_\mu p_\nu-u_\nu p_\mu
+\frac{q}{2m_e}(F_{\mu\alpha}S^\alpha\,_\nu-F_{\nu\alpha}S^\alpha\,_\mu).\label{6}
\end{equation}
Note the units with $c=1$ and the two coupling constants: charge $q$ and charge to mass ratio
$q/m_e$.

The equations admit an integral of the motion,
\begin{equation}
 m=p^\mu u_\mu=m_e+\frac{q}{2m_e} F_{\alpha\beta}S^{\beta\alpha},\label{7}
\end{equation}
which shows that the mass is a dynamical quantity and specifies a shift in mass due to interaction.

\section{Mass, frequency and time scaling}

For purposes of measurement, we must relate electron time to some observable time scale.
A suitable relation is defined by the momentum vector $ p_\mu=m v_\mu$, which determines
the velocity $v_\mu=dx_\mu/d\tau$ for a timelike history $x_\mu=x_\mu(\tau)$, and thus
associates a proper time with the electron motion.

The zitter model can now be simplified and clarified by introducing a unit \textit{zitter
vector}\,\,$e_\mu$ defined by
\begin{equation}
\omega_Z e_\mu =\frac{4}{\hbar^2} p^\nu S_{\nu\mu},\label{8}
\end{equation}
where
\begin{equation}
\omega_Z \equiv \frac{2m}{\hbar} =\omega_e+\frac{q}{2\omega_e}
F_{\alpha\beta}S^{\beta\alpha}
\equiv \frac{d\varphi}{d\tau}\label{8a}
\end{equation}
can be identified as  a \textit{zitter frequency}
and defines a \textit{zitter phase angle} $\varphi$, which is closely
analogous to the QM phase angle in the Dirac equation.

It follows that $p^\mu e_\mu=0$, so we can reduce the spin tensor to the
simple form
\begin{equation}
 S_{\mu\nu}=\frac{\hbar}{2}(u_\mu e_\nu-e_\mu u_\nu).\label{9}
\end{equation}
Moreover, the particle equation (\ref{4}) is reduced to
\begin{equation}
 \frac{d u_\mu}{d\tau}  =\omega_Z e_\mu+qF_{\mu\nu}u^\nu.\label{10}
\end{equation}
As suggested by this equation and confirmed by further analysis, the zitter vector $e_\mu$ rotates
temporally with proper frequency $\omega_Z$, so the electron's history can be described as a
lightlike helix winding about a timelike curve determined by the momentum. The variable
radius of the helix is inverse to the frequency.  According to (\ref{7}), the zitter
frequency reduces to the $2\omega_B$ for a free particle and corrections to this are
generally small, so the radius of the zitter is on the order of a Compton wavelength.

The momentum equation (\ref{5}) can be put in the more convenient form
\begin{equation}
 \frac{dp_\mu}{d\tau}=qF_{\mu\nu}u^\nu +\partial_\mu\Phi,\label{11}
\end{equation}
where the \textit{spin potential} $\Phi$ is given by
\begin{eqnarray}
\Phi =\frac{q}{2m_e}F^{\alpha\beta}S_{\beta\alpha} &=&
q\lambda_e F^{\alpha\beta}e_\alpha u_\beta\notag\\
\hbox{with}\qquad\lambda_e&\equiv&\frac{\hbar}{2m_e}=\frac{1}{\omega_e}.\label{12}
\end{eqnarray}
According to (\ref{8a}) the spin potential doubles as a frequency shift.
In general the momentum will \textit{wobble} about a mean value driven by zitter on the
right side of (\ref{12}), but that will be irrelevant to our application to experiment.

To factor out effects of zitter on the motion, we average over a zitter period holding
$v_\mu$ fixed. Denoting zitter means by an overline, we have
\begin{equation}
\obv_\mu=v_\mu=\obu_\mu,\qquad \hbox{and}\qquad\obe_\mu=0,\label{13}
\end{equation}
so, if $F_{\mu\nu}$ is not resonant with zitter fluctuations, equation  (\ref{11}) reduces to
\begin{equation}
\frac{d(\obm v_\mu)}{d\tau}=q\obF_{\mu\nu}v^\nu +\partial_\mu\overline{\Phi}.\label{14}
\end{equation}
This is the classical equation for a structured charged particle subject to a Stern-Gerlach
force with potential
\begin{eqnarray}
\overline{\Phi}=\frac{q}{2m_e}F^{\beta\alpha}&\obS_{\alpha\beta}&\notag\\
\hbox{where}\qquad&\obS_{\alpha\beta}&=\epsilon_{\alpha\beta\mu\nu}s^\mu v^\nu\label{15}
\end{eqnarray}
defines a \textit{spin vector} $s_\mu$ with constant magnitude $s_\mu s^\mu=(\hbar/2)^2$ and
 $s_\mu v^\mu=0$.

Finally, we can split the spin tensor into a fluctuating zitter component and a slowly
precessing spin component:
\begin{equation}
S_{\alpha\beta}=\frac{\hbar}{2}(v_\alpha e_\beta-v_\beta e_\alpha)+
\obS_{\alpha\beta}.\label{16}
\end{equation}
This determines a corresponding split in the spin potential:
\begin{equation}
\Phi=q\lambda_e F^{\alpha\beta}v_\alpha e_\beta+\overline{\Phi}.\label{17}
\end{equation}

For physical interpretation, it is convenient to express $\Phi$ in terms of electric and
magnetic fields $\E_v$ and $\B_v$ as seen in the \textit{rest system} of the electron
defined by the instantaneous direction of $v_\mu$. Thus,
\begin{equation}
\Phi=-\d\bdot \E_v -\bmu \bdot \B_v.\label{18}
\end{equation}
The last term on the right is the familiar Zeeman interaction energy, where $\bmu=(q/m_e)\s$
is the electron magnetic moment.
The first term is the interaction energy of an electric dipole with $\d \bdot \bmu=0$.
The \textit{zitter dipole moment} $\d=q\lambda_e \e$ rotates in the plane orthogonal to $\s$ with the
zitter frequency $\omega_Z=2m/\hbar$.
This fluctuating electric dipole interaction is here proposed
as a physical mechanism to explain the channeling resonance.

\section{ Motion along crystal channels}

For a quantitative account of electron channeling, interaction with the crystal is modeled by
a static electric potential $V=V(\x)$ in the lab frame, so $q\E =-\bnabla V$.
From the time component of equation (\ref{11}) we immediately get the familiar energy
conservation law
\begin{equation}
E=p_0 +V=\obp_0 +V, \label{19}
\end{equation}
where we note that the kinetic energy term must be equivalent to its zitter average.

The spatial component of equation (\ref{11}) can put in the form
\begin{equation}
 \frac{d\p}{dt}=-\bnabla( V+\gamma^{-1}\Phi),\label{20}
\end{equation}
where the time dilation factor $\gamma \approx u_0$ has been factored out.
In the channeling domain the maximum crystal potential is a few hundred electron volts at most, so
in the 80 MeV region of interest to us, the effective electron mass $M\equiv\gamma m_e =E/c^2$ is
constant to an accuracy of $10^{-5}$, and $\gamma=158$.

In axial channeling electrons are trapped in orbits spiraling around a crystal axis. To a first
approximation, the crystal potential can be modeled as the potential for a chain of atoms, so it
has the form
\begin{equation}
 V(\x)=V(r,z)=U(r)P(2\pi z/d) ,\label{21}
\end{equation}
where $\x(t)=\r+z\bsig_z$ is the particle position from the first atom in the chain, with
$r=|\r|$. The longitudinal potential $P(2\pi z/d)= P(\omega_0 t)$ is periodic with a \textit{ tunable
frequency} $\omega_0=2\pi\dot{z}/d$ that varies with the energy $E$.

Our problem is to calculate perturbations on the transverse component of the momentum
vector, as that can remove electrons from stable orbits in the beam. The transverse
component of equation (\ref{20}) has the familiar form of a nonrelativistic equation:
\begin{equation}
 M\frac{d^2\r}{dt^2}=-\hat{\r}V'-
\gamma^{-1}(\bnabla \Phi)_\perp,\label{22}
\end{equation}
where $V' =\partial_r V=U'P$. Ignoring the zitter perturbation for the time being, we seek to
ascertain the effect of the periodic factor $P(\omega_0t)$ on the orbital motion. Oscillations
in the transverse velocity can be ignored.

\subsection{Rosette channeling with atomic perturbations}

For analytic simplicity, we approximate the
potential  by the first two terms in a Fourier expansion with respect to the
reciprocal lattice vector:
\begin{equation}
V=U(r)(1+\cos\omega_0t) ,\label{23a}
\end{equation}
with
\begin{equation}
U(r)=-k\ln [1+(Ca/r)^2],\label{23}
\end{equation}
where the first term is Lindhard's potential for a uniformly charged
string \cite{Gemmell74,Lindhard65}, and the coefficient of the second term is set to make the
potential vanish between atoms. The parameter $a=0.190\AA$ is the Fermi-Thomas
screening radius, and the constant $C^2=3$ is a fairly accurate fit over the range of
interest. For silicon ($Z=14$) the coupling constant has the value $k= Ze^2/d =52.5$ eV.
In most channeling applications the second term is ignored; however, its periodicity is essential
for resonance in the zitter perturbation investigated here.

It will be convenient to represent the radius vector in the complex form
$\r=re^{i\theta}$, where the imaginary $i$ is generator of rotations in the transverse plane.
Then equation (\ref{22}) assumes the complex form.
\begin{equation}
 M\frac{d^2\r}{dt^2}=-U'Pe^{i\theta}=-\frac{U'}{r}(1+\cos\omega_0t)\r.\label{24}
\end{equation}
We are interested only in radial oscillations, so we use conservation of angular
momentum $L=Mr^2\dot{\theta}$ to separate out the rotational motion.
With the periodic driving factor omitted for the moment, equation (\ref{24}) admits the energy
integral
\begin{equation}
 E_\perp =\frac{1}{2}M\dot{r}^2 + W(r).\label{24a}
\end{equation}
where
\begin{equation}
W(r) =\frac{L^2}{2Mr^2} +U(r).\label{24b}
\end{equation}Let us expand this around a circular orbit of radius $r_0$, and for quantitative estimates take
$r_0=0.50 \AA$ as a representative intermediate radius. For $\r_0=re^{i\theta_0}$ and
$U_0'=U'(r_0)$, equation (\ref{24}) gives us
\begin{equation}
\dot{\theta_0}^2=\frac{L^2}{M^2r_0^4}=\frac{U_0'}{Mr_0}=\frac{ 31.7}
{80.9\times10^6}
\left(\frac{c^2}{r_0^2}\right).\label{25}
\end{equation}
In terms of $x=r-r_0$, second order expansion of (\ref{24a}) gives us
\begin{equation}
 E_\perp =\frac{1}{2}M[\dot{x}^2  +\Omega_0^2x^2]+ W_0,\label{26}
\end{equation}
where, for mass $M$ at the expected resonance, we have
\begin{align}
\hspace{-1.4cm}\Omega_0^2\equiv \frac{W_0''}{M}&=\frac{3U_0'+r_0U_0''}{Mr_0}\notag\\
&=\frac{3\times 31.7-76.0}
{80.9\times10^6}\left(\frac{c^2}{r_0^2}\right),\label{29}
\end{align}
Thus, we have $\Omega_0=4.21\times10^{15}\,\rm{s}^{-1}$, which
is close to
$\dot{\theta_0}=4.75\times10^{15}\,\rm{s}^{-1}$ from (\ref{25})
and much less than the expected resonant frequency $\omega_0=\omega_B/\gamma=4.91\times
10^{18}\,\rm{s}^{-1}$. Note that the distance traveled in one orbital revolution is
$d_r=2\pi c/\dot{\theta_0}=3.97\times 10^3 \AA=0.397\rm{\mu m}$. Thus, the orbit makes 2.52
revolutions in passing through the one micron crystal, so the orbital
revolutions are only weakly coupled to the high frequency radial oscillations.

Now differentiating  (\ref{26}) and reinserting the periodic driving
factor, we obtain the desired equation for radial oscillations:
\begin{equation}
\ddot{x} +\Omega_0^2(1+\cos\omega_0t)x=0 .\label{28}
\end{equation}
This equation has solutions of the general form
\begin{equation}
x(t)=e^{i\Omega t}\sum_{n=-\infty}^{\infty}a_ne^{in\omega_0 t}.\label{31}
\end{equation}
Since $\Omega_0 << \omega_0$ the solution is well approximated by first order terms, which yield
the particular solution
\begin{equation}
x(t)=a(\cos\Omega t)\cos\omega_0t=\frac{a}{2}(\cos\omega_+ t+\cos\omega_- t),
\label{33}
\end{equation}
\begin{equation}
\hbox{with} \quad\Omega^2=\frac{3}{2}\Omega_0^2 \quad \hbox{and} \quad
\omega_\pm =\omega_0\pm \Omega.\label{32}
\end{equation}
This describes a harmonic oscillator with high frequency
$\omega_0$ and a slowly varying amplitude with frequency $\Omega$, which is equivalent to a sum
of two oscillators with frequencies $\omega_{\pm}$ separated by $\omega_+ -\omega_-=2\Omega$.

We shall see that, at the resonant frequency $\omega_0=\omega_B/\gamma$, the frequency shift
$\Omega=(0.857\times 10^{-3})\omega_0$ is the right order of magnitude to contribute to
experimental effects. Moreover, this quantity has been estimated at the particular radius
$r=0.50 \AA$, and it may be larger by an order of magnitude for smaller radii of experimental
relevance. Accordingly, a distribution of $\Omega$ values will contribute to the experiment.

\subsection{Zitter resonance}
Now we are prepared to consider the effect of zitter perturbations on the orbit.
Inserting the crystal electric field with its relativistic transformation into (\ref{18}) and
keeping only the largest term, it can be shown that \cite{Hest08}
\begin{equation}
\Phi=-\lambda_e\gamma U'P\e\bdot \hat{\r} =-\lambda_e\gamma U'P\cos(\omega_Z t/\gamma+\delta)
.\label{55}
\end{equation}
 By the way, we don't need to use the spin equation of motion
(\ref{6}) in our calculations; we only need the fact that it implies the unit zitter vector
$\e$ rotates rapidly in a plane that precesses slowly with to the spin vector $\s$. We have dropped
the spin contribution to $\Phi$, because a more elaborate experiment would be needed to detect it.
Also, according to equation (\ref{8a}), the spin potential (\ref{55}) contributes to an oscillating
shift in zitter frequency, but that effect is too small to be detected in the present experiment \cite{Hest08}, so in the rest of our analysis we suppose $\omega_Z=\omega_e$.

Inserting (\ref{55}) into  (\ref{22}) with a convenient choice of phase and writing
$\omega=\omega_Z/\gamma$, we get the
 equation of motion
\begin{equation}
 M\frac{d^2\r}{dt^2}=-\r \frac{U'}{r}(1+\cos\omega_0 t)(1+\frac{\lambda_e}{R}
\cos\omega t),\label{58}
\end{equation}
where $R\equiv -U'/U''=r[1 +(Ca/r)^2]/[3 +(Ca/r)^2]$ is an \textit{effective  radius}.
Ignoring the amplitude modulation as determined in (\ref{33}), we can reduce this to a radial
equation
\begin{equation}
 \ddot{x}+\omega_0^2(1+\frac{\lambda_e}{R}
\cos\omega t)x=0.\label{59}
\end{equation}
Of course, we can replace $\omega_0$ in this equation by $\omega_{\pm}$ to get two separate
resonant peaks.

For $h=\lambda_e/R$ constant, (\ref{59}) is Mathieu's equation, for which there are well known
methods of solution. So let us solve it using
$h=\lambda_e/R_0=1.931\times10^{-3}/0.208=9.283\times 10^{-3}$ as an approximation. For small
values of $h$ such as this, it can be shown that (\ref{59}) has resonances at $2\omega_0=n\omega$,
for
$n=1,2,.\,.\,.$.  We demonstrate that for the first order resonance by assuming a solution of the form \cite{Landau69}
\begin{equation}
x=e^{st}[a\cos\frac{\omega t}{2}+b\sin\frac{\omega t}{2}],\label{60}
\end{equation}
where the so-called Floquet exponent $s$ must be real and positive.

Looking for solutions with $\omega =2\omega_0+\epsilon$ close to the supposed first order
resonance, we insert (\ref{60}) into the differential equation (\ref{59}) and keep only first order
terms to derive the conditions
\begin{equation}
  sb-\frac{1}{2}[\epsilon-\frac{1}{2}h\omega_0]a=0,\quad
sa+\frac{1}{2}[\epsilon+\frac{1}{2}h\omega_0]b=0.\label{61}
\end{equation}
These equations can be solved for the coefficients provided
\begin{equation}
 s^2=\frac{1}{4}[(\frac{1}{2}h\omega_0)^2-\epsilon^2]>0.\label{61a}
\end{equation}
Thus, we do indeed have a resonance with width
\begin{equation}
\Delta \omega=2\epsilon=h\omega_0=9.282\times10^{-3}\omega_0,\label{62}
\end{equation}
or
\begin{equation}
\Delta p=h(80.9\,\rm{MeV/c})=0.751\,\rm{MeV/c}.\label{63}
\end{equation}
And for the amplification factor at resonance we have
\begin{equation}
st=\frac{1}{4}h\omega_0t=\frac{\pi}{2}hn=1.46\times10^{-2}n,
\label{64}
\end{equation}
where $n$ is the number of atoms traversed in a resonant state.
Since $\ln2=0.693$, this implies that the
amplitude is doubled in traversing about $50$ atoms.

Now note that the value of $h=\lambda_e/R_0$ used in (\ref{62}) and (\ref{64}) applies to only the
subclass of orbits for which $r_0=0.50 \AA$. For smaller radii the values can be much larger.
In principle, to get the width of the ensemble of orbits we should replace $h$ in (\ref{62}) by
its average $\obh$ over the ensemble. However, the result will probably not differ much from the
typical value we have chosen.

Similarly, the amplitude factor in (\ref{64}) will have a distribution of values, and the
doubling factor will be reached much faster for orbits with smaller radii.
Presumably, random perturbations (such as thermal fluctuations of the nuclei) will limit the
resonant state coherence length to some mean value $\overline{n}$. Consequently, states
with smaller $r_0$ will be preferentially ejected from the beam.

 Even more to the point, the perturbation parameter $h=\lambda_e/R$ is not constant as we
tentatively supposed but increases rapidly as the electron approaches a nucleus. In resonance
the value of $h$ close to each nucleus dominates the perturbation, so its effective mean value
is much smaller than the estimate for constant $r_0$.  Evidently, resonant interaction may
eject electrons with small $r_0$ in just a few atomic steps.

\section{ Discussion and Conclusions}

The predicted resonance width in (\ref{63}) is in fair agreement with the width in the
channeling experiment data \cite{Gouanere05}, considering uncertainties in the value of $h$ and
such factors as thermal vibrations that may contribute to damping. Damping can only
narrow the width and destroy the resonance if it is too severe.

We need to explain how the orbital resonance is manifested in the experimental
measurements.   Two scintillators, SC2 and SC3, were employed to detect
the transmitted electrons. The larger detector SC3, with a radius about 3 times that of SC2,
served as a monitor while the smaller detector served as a counter for a central portion of
the beam. The measured quantity was the ratio of SC2 to SC3 counts as momentum was varied
in small steps $(0.083\%)$ over a 2\% range centered at the expected resonance momentum
$80.9$ MeV. An 8\% dip was observed at $81.1$ MeV.
Orbital resonances may contribute to this effect in at least two ways: first, and perhaps
most important, by increasing the probability of close encounter with a nucleus that will
scatter the electron out of the beam; second, by increasing the duration of eccentric orbits
outside the central region.

For the most part, resonant interactions will be strongest on electrons
confined to the central region.
However, it should be noted that in silicon each $<110>$ channel has a nearest neighbor chain
only $1.36\AA$ away with the location of atoms shifted by precisely half an atomic step
$1.92\AA$.
Considering the slow precession of a channeled orbit at the resonant frequency, this
chain will resonate with it for hundreds of atomic steps. In fact, the interaction might lock
onto the orbit to prevent precession during resonance. This could indeed have a substantial
effect on resonant channeled electrons, so it deserves further study.
These observations suffice for a qualitative explanation of the observed dip.
A quantitative calculation will not be attempted here.
Indeed, for a fully quantitative treatment the string potential (\ref{23}) must be replaced by a
more realistic crystal potential, for, as Lindhard \cite{Lindhard65} has shown, it breaks down at
atomic distances near the screening radius $a=0.190\AA$.
Even so, the string approximation is useful for
semiquantitative analysis of zitter resonance, as we have seen.

The most problematic feature of the experiment is the $0.226\,\rm{MeV/c}$ difference
between observed and predicted resonance energies. If estimation of the experimental error was
overly pessimistic, that indicates a physical frequency shift. The most likely origin for such a
shift is the frequency split in (\ref{33}).  The experiment was not sufficiently accurate to
resolve separate peaks for the two frequencies, so the peaks would merge to broaden the
measured resonance width.  However, the peak for $\omega_+$ is likely to be higher than the
peak for $\omega_-$ owing to greater probability for ejection from the beam. Hence, the
center of the merged distribution will be displaced to a higher frequency. If this explanation is
correct, then an increase in experimental resolution will separate the two peaks, and their
relative heights will measure the relative probability of ejection at the two frequencies.

If the idea of zitter resonance is taken seriously, there are many opportunities for new
theoretical and experimental investigations.  Increasing the resolution by three orders of
magnitude will open the door to refined studies of frequency shifts, line splitting, spin
effects and Zeeman splitting, all of which are inherent in zitter theory \cite{Hest08}.
As has been noted, the most straightforward prediction of the zitter model is a second
order resonance near $161.7 \rm{MeV/c}$. In a first approximation, it can be analyzed in
much the same way as here, though removal of electrons from the center of the beam may be
enhanced by such processes as pair creation.

Classical particle models have long been used for channeling calculations with considerable
success. Besides being simpler and more transparent than quantum mechanical models, they
even give better results at high energies \cite{Gemmell74}.
For present purposes, the zitter model differs from the usual classical model only by the
zitter perturbation in equation (\ref{20}). It is unlikely that the zitter perturbation effect
can be explained by standard quantum theory, though it might be by a subtle modification of the
Dirac equation \cite{Hest08}. The zitter model
has many other physical implications \cite{Hest08}, though channeling is so far the most direct
way to test it.

All things considered, we conclude that the quasiclassical zitter model provides a
plausible explanation for the reported experimental data. This is ample reason to refine and
repeat the experiment to confirm existence of the resonance.

\begin{acknowledgments}
I am deeply indebted to Michel Gouan\`ere for
explaining details about his channeling experiment and data analysis.
\end{acknowledgments}

\bibliography{zbw}

\end{document}